\documentclass[12pt,preprint]{aastex}
\usepackage{rotating}
\renewcommand{\cite}{\citealp}

\def\omc{\hbox{$\omega$~Cen~}}
\def\omcp{\hbox{$\omega$~Cen}}
\def\hhf{\hbox{\em hot helium flashers}} 
\def\rr{\hbox{RR Lyrae~}}

\begin{document}

\title{On the period distribution of cluster RR Lyrae stars to constrain 
their helium content: the case of $\omega$ Centauri}

\author{M. Marconi\altaffilmark{1}, 
G. Bono\altaffilmark{2,3,4}, 
F. Caputo\altaffilmark{3}, 
A. M. Piersimoni\altaffilmark{5},  
A., Pietrinferni\altaffilmark{5}, 
R., F. Stellingwerf\altaffilmark{6}}   

\altaffiltext{1}{INAF-Osservatorio astronomico di Capodimonte, Via Moiariello 16, 80131 Napoli, Italy; marcella.marconi@oacn.inaf.it}
\altaffiltext{2}{Dipartimento di Fisica - Universit\`a di Roma Tor Vergata, Via della Ricerca Scientifica 1; giuseppe.bono@roma2.infn.it}
\altaffiltext{3}{INAF-Osservatorio Astronomico di Roma, Via Frascati 33, 00040 Monte Porzio Catone, Italy; caputo@oa-roma.inaf.it}
\altaffiltext{4}{European Southern Observatory, Karl-Schwarzschild-Str. 2, 85748 Garching bei Munchen, Germany}
\altaffiltext{5}{INAF-Osservatorio Astronomico di Collurania, Via M. Maggini, Teramo, Italy; piersimoni@oa-teramo.inaf.it, adriano@oa-teramo.inaf.it}
\altaffiltext{6}{Stellingwerf Consulting, 11033 Mathis Mtn Rd SE, 35803 Huntsville, AL USA; rfs@swcp.com}

\begin{abstract}

We present new sets of nonlinear, time-dependent convective hydrodynamical 
models of \rr stars assuming two metal (Z=0.0005, Z=0.001) and 
three helium abundances (Y=0.24, 0.30, 0.38). For each chemical composition we 
constructed a grid of fundamental (FU) and first overtone (FO) models 
covering a broad range of stellar masses and luminosities. 
To constrain the impact of the helium  content on \rr properties, 
we adopted two observables --period distribution, luminosity 
amplitudes-- that are independent of distance and reddening. The current 
predictions confirm that the helium content has a marginal effect on the 
pulsation properties. The key parameter causing the difference between 
canonical and He-enhanced observables is the luminosity.                                   
We compared current predictions with the sample of 189 \rr stars in 
\omc and we found that the period range of He-enhanced models 
is systematically longer than observed.  These findings apply to 
metal-poor and metal-intermediate He-enhanced models.
To further constrain the impact of He-enhanced structures on the period distribution 
we also computed a series of synthetic HB models and we found that the predicted period 
distribution, based on a Gaussian sampling in mass, agrees quite well with observations. 
This applies not only to the minimum fundamentalized period of RR Lyrae stars 
(0.39 vs 0.34 day), but also to the fraction of Type II Cepheids (2\% vs 3\%). 
We also computed a series of synthetic HB models assuming a mixed HB population in 
which  the 80\%  is made of canonical HB structures, while the 20\% is made of 
He-enhanced (Y=0.30) HB structures. We found that the fraction of Type II Cepheids 
predicted by these models is almost a factor of two larger than observed 
(5\%  vs 3\%). This indicates that the fraction of He-enhanced structures in 
\omc cannot be larger than 20\%.  
\end{abstract}


\section{Introduction} 

The stellar content of globular clusters (GCs) considered an optimal realization of 
a simple stellar population, since they are coeval and chemically homogeneous stellar 
systems, has been questioned. 
The spectroscopic evidence of star-to-star variations in the abundance of 
C, N and Na as well as of Al and O in many cluster Red Giant (RG) stars 
dates back to forty years ago \citep{Osborn71}.    
Moreover, the molecular band-strengths of CN and CH seem to be anti-correlated 
\citep{smith87,Kraft94} and anti-correlations between O--Na and Mg--Al have been 
observed in evolved (RG, Horizontal Branch [HB]), and in unevolved (Main Sequence [MS]) 
stars of all the GCs investigated with high-resolution spectra 
\citep{pila83,grat04}. 

More recently, accurate and deep Hubble Space Telescope optical photometry disclosed 
the presence of multiple stellar populations in several massive GCs. Together with the 
most massive Galactic GC \omc \citep{ande02,be04} multiple stellar sequences 
have been detected in GCs covering a broad range of metal content: NGC~2808 
\citep{piot07}, M54 \citep{sieg07}, NGC~1851 \citep{cala07,milo08} and 47 Tuc 
\citep{dicri10,nata11}. 
Some of these multiple sequences (\omcp, NGC~2808, NGC~1851) might be explained 
either with a He-enhanced\citep{norr04,lee05,dant08,piot07}, or with a CNO-enhanced 
\citep{cala07,cass08} sub-population. However, no general consensus has been reached 
concerning the physical mechanisms, the evolutionary history and the fraction of these 
stellar components \citep{berst09}.  

The possible occurrence of He-enriched stars in GCs was suggested   
to explain not only the presence of multiple unevolved sequences, but also the 
presence of extended blue HB tails \citep[][and references therein]{dant08}. 
Evolutionary prescriptions indicate that He-enriched structures, at fixed 
metallicity and cluster age, have a smaller turn-off mass when compared with structures
constructed by assuming a canonical He content. Therefore, a He-enriched sub-population,
for a fixed mass loss rate, is characterized by smaller envelope masses and will 
{\em mainly} populate the hot and the extreme region of the HB \citep{bo10}. 
Therefore, the presence of He-enriched sub-population(s) in GCs alleviates the 
heavy assumption of strong stochastic changes in the mass-loss rate along the RGB, 
required by the canonical scenario to populate the entire HB.

In this context \omc plays a key role, since it is a massive GC \citep[][]{lee09} 
in which have been identified several MS and sub-giant branches 
\citep[][and references therein]{bel10}. Low-resolution spectra indicate that the bluest MS 
is more metal-rich than the redder ones, thus further supporting the evidence that it could 
be He-enhanced \citep{norr04,pio05}. However, Castellani et al. (2007) using star count 
ratio of HB and RG stars suggested that a significant fraction of Extreme HB (EHB) stars in  
\omc might be aftermath of the \hhf~ 
scenario\footnote{Evolutionary models indicate that HB structures which ignite He  
along the white dwarf cooling sequence (\hhf) experience a significant mixing 
\citep{Miller08,Brown10} between the core (He and carbon-rich) and the envelope 
(hydrogen rich). The structures experiencing this flash-mixing phenomenon show an 
increase in the surface abundance of carbon ranging from 1\% to 5\%  and an increase 
in He content. This is a crucial observable, since the progeny of the He-enriched 
scenario is not supposed to be carbon-enriched.}.  
By using a broad range of synthetic color-magnitude diagrams (CMDs) \citet[][]{cas09} 
found that EHB stars ($Te \ge 30,000$ K) in \omc should be a mix of He-enhanced, 
\hhf~ and canonical HB stars.  
In a spectroscopic investigation of \omc EHB stars, \citet{moeh10}
found that roughly the 30\% are He-poor, while the 70\% have either solar 
or super-solar He abundances. Note that the carbon-rich stars are also He-rich stars, 
thus further supporting the evidence that a fraction of these stars is the progeny 
of the \hhf~ scenario. 

In a recent evolutionary investigation \citep[][]{dan10} suggest that very high 
He abundances (Y=0.80) could be required to explain \omc EHB stars and 
that warm ($log T_e$$<$10,000 K, see their Fig.~2) HB stars  might have 
He abundances ranging from the canonical to significantly higher He abundances. 
However, \citet{s06} using  high-resolution spectra of 74 \rr stars in \omc 
found that their metallicity distribution agrees quite well with the metallicity 
distribution of RG stars \citep{cala09}.
Moreover, the distribution in the CMD of metal-intermediate \rr stars agrees quite well 
with the predicted metal-intermediate Zero-Age Horizontal Branch (ZAHB) based on 
HB models constructed assuming a canonical 
He content. The quoted authors pointed out that if the blue MS is made of He-enhanced, 
metal-intermediate MS stars, and if these structures are the progenitors of the EHB stars, then 
\omc should simultaneously host two metal-intermediate stellar populations with two 
different He contents.

In this investigation we address the helium content of \omc \rr using their period 
distribution and luminosity amplitude. 

\section{The theoretical framework}
The new pulsation \rr models have been computed by using the hydrodynamical 
code developed by \citet{s82} and updated by \citet[][see also \citet{smolec10} 
for a similar approach]{bs94,bms99}. The physical assumptions adopted to compute 
these models will be described in Marconi et al. (2011, in preparation). We adopted 
the OPAL radiative opacities released in 2005 by 
\citep[][http://www-phys.llnl.gov/Research/OPAL/opal.html]{ir96} 
and the molecular opacities by \citet{af94}.  
To constrain the pulsation properties of \rr stars in \omcp, we adopted two 
metal abundances similar to the two observed peaks in the \rr metallicity distribution 
\citep[][]{s06}, namely Z=0.0005 and Z=0.001. For each metal content we adopted three He  
contents to constrain the dependence of the pulsation properties of \rr stars on this 
crucial parameter. For each fixed chemical composition the stellar mass of \rr stars 
was fixed by using the evolutionary prescription for $\alpha$-enhanced structures provided by 
\citet{piet06} and available on the BASTI database\footnote{http://albione.oa-teramo.inaf.it/}. 
Note that for each fixed metal content the mass of the He-enhanced models was estimated 
assuming the same cluster age (13 Gyr).  
Together with the luminosity predicted by evolutionary models we often adopted a brighter 
luminosity level to account for the possible occurrence of evolved 
\rr stars. The input of the different sets of He-enhanced models are listed in Table 1, 
where columns from 1 to 4 give the metal and the He abundance, the stellar mass and the 
luminosity. 

\begin{deluxetable}{llll}
\tablewidth{0pt}
\tablecaption{Intrinsic stellar parameters adopted to compute 
canonical and He-enriched pulsation models.}
\tablehead{
\multicolumn{1}{c}{Z\tablenotemark{a}} &
\multicolumn{1}{c}{$Y$\tablenotemark{a}} &
\multicolumn{1}{c}{$M/M_{\odot}$\tablenotemark{b}} &
\multicolumn{1}{c}{$\log L/L_{\odot}$\tablenotemark{c}} 
}
\startdata
0.0004& 0.24\tablenotemark{c} & 0.70 & 1.6, 1.7, 1.8  \\
0.0005& 0.35 & 0.65 & 1.85, 1.95 \\
      & 0.40 & 0.60 & 1.9 \\
0.001 & 0.24\tablenotemark{c} & 0.65 & 1.5, 1.6, 1.7, 1.9 \\
      & 0.30 & 0.60, 0.65 & 1.8, 1.9, 2.0 \\
      & 0.38 & 0.60, 0.65 & 1.8, 1.9, 2.0 
\enddata
\tablenotetext{a}{Metal (Z) and helium (Y) abundance by mass.}
\tablenotetext{b}{Stellar mass (solar units).}
\tablenotetext{c}{Logarithmic luminosity (solar units).}
\tablenotetext{d}{Canonical \rr pulsation models computed by 
\citet{b03,dmc04}. The difference in 
metal content between the canonical and the He-enhanced models has 
a minimal impact on the pulsation properties and they are treated 
together.}
\end{deluxetable}

\begin{figure}
\includegraphics[height=0.5\textheight,width=0.5\textwidth]{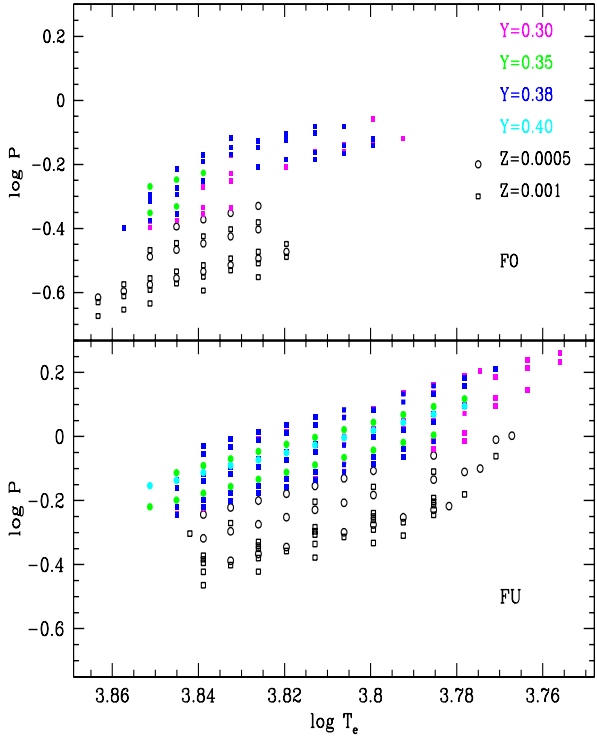}
\label{plotP}
\caption{Top -- Period distribution of metal-poor (Z=0.0005, open circles) and metal-intermediate 
(Z-0.001, squares) first overtone \rr models. The models with canonical He content are plotted 
as empty symbols, while the He-enhanced ones with filled symbols and different colors (see labels). 
Bottom -- Same as the top, but for the fundamental \rr models.}
\end{figure}

For each fixed chemical composition, mass value and luminosity level, we investigated the limit cycle 
behavior of both fundamental (FU) and first overtone (FO) pulsators by covering a wide range of effective 
temperatures. 
For each model the pulsation equations were integrated in time till the radial motions approached their 
limit cycle stability. This means that we can provide robust constraints not only on the topology of the 
instability strip (IS), but also on the pulsation amplitudes of He-enhanced models.  
The top and the bottom panel of Fig. 1 show the period distribution of canonical\footnote{The reader 
interested in a detailed discussion concerning the canonical \rr models is referred to \citet{b03,dmc04}} 
(empty symbols) and He-enhanced (filled symbols) models. The circles and the squares display metal-poor 
and metal-intermediate \rr models. 

\begin{figure}
\includegraphics[height=0.5\textheight,width=0.5\textwidth]{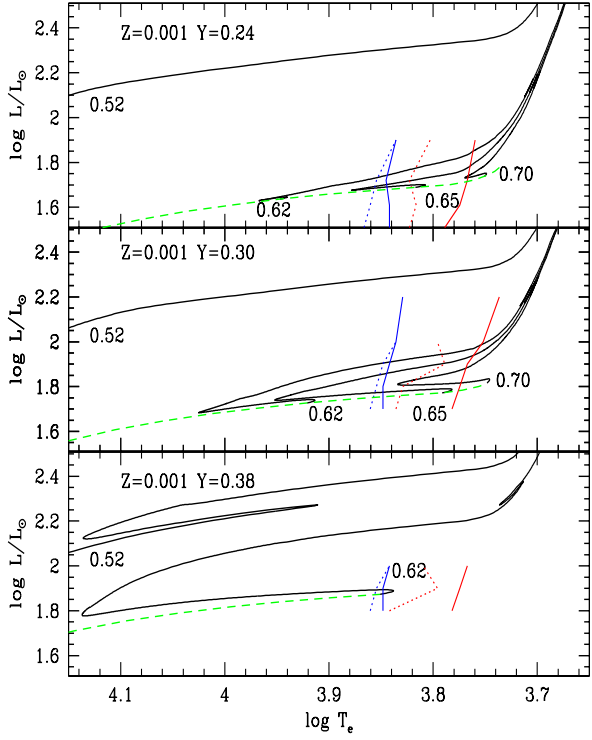}
\label{plotP}
\caption{Top -- Predicted \rr instability strip at fixed metal content (Z=0.001) based on canonical
pulsation models.  The solid and the dotted vertical lines display the instability strip for fundamental 
and first overtone RR Lyrae. The almost horizontal dashed line shows the predicted Zero-Age Horizontal 
Branch, while the solid lines selected HB evolutionary models (BASTI database) of different stellar 
masses. 
Middle -- Same as the top, but for moderately He-enhanced models (Y=0.30).
Bottom -- Same as the top, but for He-enhanced models (Y=0.38).}
\end{figure}

A glance at the data plotted in Fig.~1, indicate that \rr models with a He-enhanced composition 
show periods that are systematically longer than the canonical ones. The difference is mainly caused by 
evolutionary effects. 
The top (canonical), the middle (He-enhanced, Y=0.30) and the bottom (He-enhanced, Y=0.38) panel of 
Fig.~2 show together with the modal stability of FU (solid vertical lines) and FO (dotted vertical 
lines) \rr stars the predicted ZAHBs (dashed line) and selected HB evolutionary models.     
On the basis of the evolutionary predictions plotted in this figure we can identify two different 
regimes. 

{\em HB structures evolving inside the instability strip} --\\ 
The HB structures with a ZAHB location inside the IS and evolving during their  
central He-burning phases inside the \rr IS show, at fixed cluster age (13 Gyr) 
and  metal content (Z=0.001), similar mass values when moving from a canonical (Y=0.24, top panel 
of Fig. 2) He content to a moderately He-enhanced (Y=0.30, middle panel of Fig. 2) abundance. 
The comparison cannot be extended to the more He-enhanced models (Y=0.38, bottom panel of Fig. 2), 
since the predicted ZAHB for these models minimally intersect the predicted \rr IS. 
The mass of the progenitors decreases from 0.80 to 0.70M$_\odot$ 
and to 0.62M$_\odot$ when moving from the lowest to the highest He content. The typical masses 
populating the IS of the canonical models range from $\sim$0.64 to $\sim$0.69M$_\odot$, and the 
typical luminosity is $\log L/L_\odot \approx$1.70. For moderately He-enhanced models the masses 
range from $\sim$0.64 to $\sim$0.70M$_\odot$, and the typical luminosity is 
$\log L/L_\odot \sim$1.80.  
Evolutionary prescriptions concerning moderately He-enhanced models (Y=0.30) indicate that 
HB structures evolving inside the IS are a mix between structures evolving from the blue 
to the red and also in the opposite direction. This effect is mainly caused by the fact 
that an increase in He-content causes an extension of the excursion toward hotter 
effective temperatures that these structures perform during their off-ZAHB evolution 
(see the bottom panel of Fig.~2; Sweigart \& Gross 1976; Sweigart \& Catelan 1998; 
Fig.~2 in D'Antona et al. 2010). 

{\em HB structures evolving across the instability strip} --\\ 
The HB structures with lower total masses attain their location along the ZAHB toward effective 
temperatures hotter than \rr IS. During their off-ZAHB evolution they cross the IS 
at luminosities systematically brighter than typical \rr stars. 
For a hot HB structure of 0.52 M$_\odot$ the luminosity at the center of the IS 
($\log T_{eff}$=3.83) is $\approx$2.30 for the Y=0.24 and Y=0.30 models and becomes 
2.40 for the Y=0.38 models, while the period changes from P$\sim$2.0 to P$\sim$2.4 day.     
For a warm HB structure with a stellar mass of 0.62 M$_\odot$ the luminosity is 
$\log L/L_\odot \approx$1.76 for Y=0.24, $\approx$1.91 for Y=0.30 and 
$\approx$2.17 for Y=0.38, while the period increases  by more than a factor of two: 
P$\sim$0.61,   P$\sim$0.83 and  P$\sim$1.36 day respectively.      
The main evolutionary consequence of the presence of He-enhanced models is that 
we can have HB structures with similar masses either evolving inside the IS 
or crossing the IS that are from 0.1 to $\sim$0.4 dex more luminous than 
the canonical ones. The main 
outcome of the quoted differences is to produce \rr stars with pulsation periods 
that are systematically longer than the canonical ones.          

The above findings concerning the increase in the period of RR Lyrae stars supports 
the simulations provided by Sweigart \& Catelan (1998) using He-enhanced models.
The quoted authors investigated three different noncanonical scenarios to explain 
the tilted HB of two metal-rich GCs (NGC~6388, NGC~6441). They also investigated 
the impact on the period distribution of RR Lyrae and found that the increase 
in He causes a systematic increase in the luminosity, and in turn in the period 
of RR Lyrae stars (see their Fig.~3).  

Two evolutionary effects are noteworthy concerning He-enhanced 
HB models. {\em Mass loss}-- He-enhanced models \citep{cas09,dicri10} alleviate 
the requirement 
of strong changes in the mass loss rate along the RGB. Current predictions also 
indicate that the mass loss rate of RG stars is independent of their He 
content. This means that if we assume multiple populations with different 
He-contents the ZAHB from the red HB to the EHB will be {\em naturally} 
populated with structures of different total mass.  
{\em Lifetime}-- Evolutionary prescriptions indicate that the central He-burning 
lifetime of He-enriched structures is longer than for canonical HB \citep{cas09,dan10}. 
The difference with the most He-enriched models ranges from 
10\% for red HB stars to 30\% for EHB stars ($\log$Te$\approx$4.3). The increase in 
the lifetime is caused by the fact that they are fainter, since the He-core mass 
of He-enhanced models is slightly smaller than the canonical ones. In these structures, 
the H-shell plays a marginal role, since the envelope mass is quite small. The HB region 
in which the H-shell becomes relevant in the energy budget is for effective temperatures 
cooler than $\log$Te$\approx$4, and indeed in this region He-enhanced structures are 
brighter than canonical ones. However, the He-burning lifetime of He-enhanced models 
is longer over the entire temperature range. The difference is smaller and appears  
counter-intuitive, since the former structures are, at fixed effective temperature, 
brighter than the latter ones.
The increase in the He-burning lifetime is mainly caused by the fact that the
hydrogen-burning shell in the He-enriched structures is more efficient, due to
the increase in the mean molecular weight (see Fig.~6 in Bono 2010). This means that 
the moderately He-enhanced sub-population is expected to populate the entire HB.

\section{Comparison with observations}

\begin{figure}
\includegraphics[height=0.6\textheight,width=0.5\textwidth]{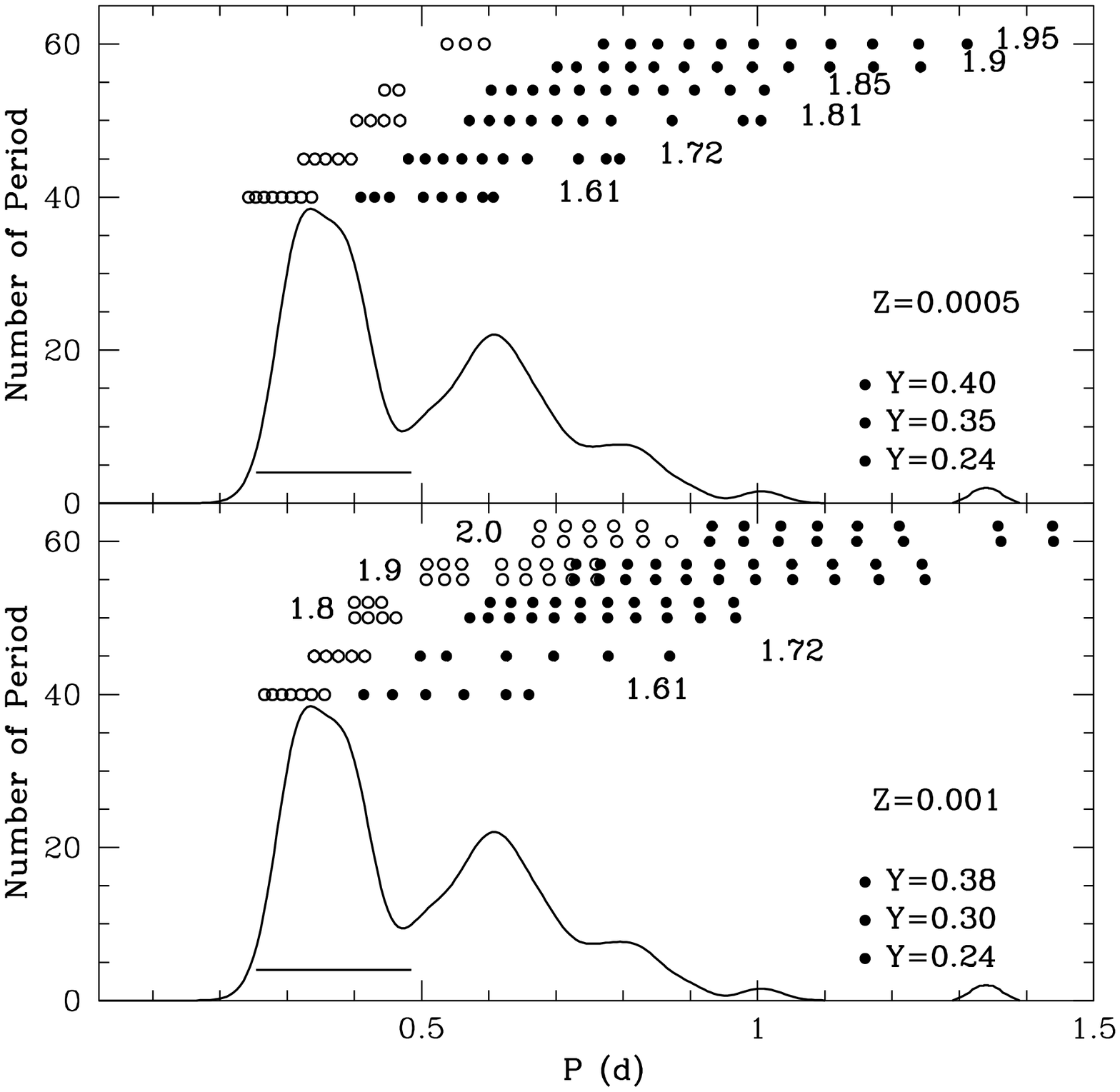}
\caption{Top -- Smoothed period of \rr stars in \omc compared with metal-poor 
($Z=0.0005$) predicted periods computed assuming canonical and He-enhanced abundances. 
The symbols are the same as in Fig.~1. The adopted He abundances and luminosity levels 
are labelled. The horizontal solid line plotted inside the observed period distribution 
shows the period range covered by FO \rr in \omcp. 
Bottom -- same as the top, but for metal-intermediate structures ($Z=0.001$).\label{fig3}}
\end{figure}

To constrain the He content of \rr stars in \omc we adopted the period distribution. 
The reasons to use this observable are threefold. {\em i)}-- \omc hosts a sizable sample 
of \rr stars \citep[189: 100 RRc, 89 RRab][]{Kal04,Wel07}. 
{\em ii)}-- The distance modulus of \omc was fixed using the K-band Period-Luminosity 
relation of \rr stars \citep{dp06} and with the tip of the RG branch \citep{bono08}. 
However, different authors are still adopting distance moduli that differ by more than 
0.3 dex \citep{cas09,dan10}. The period distribution is distance and reddening 
independent.  {\em iii)}-- The census of bright variables in this cluster is complete 
\citep{Kal04,Wel07}.  

\begin{figure*}
\includegraphics[height=0.5\textheight,width=0.90\textwidth]{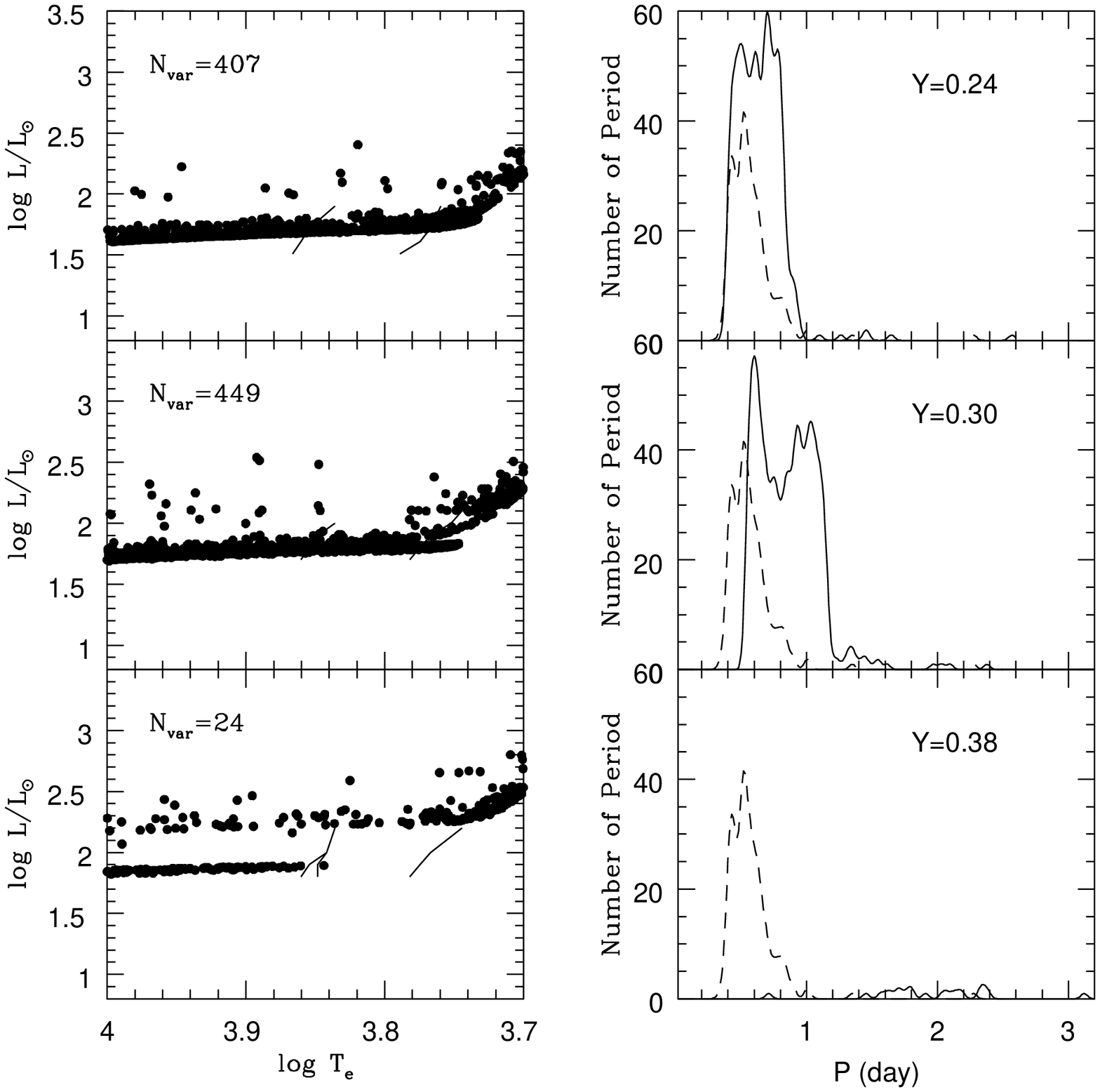}
\caption{Left -- Hertzsprung-Russell Diagram showing synthetic HB models 
constructed assuming a linear sampling in mass. The blue and the red lines
display the first overtone blue edge and the fundamental red edge. From top
to bottom the three panels display predictions based on HB models with 
different helium contents: Y=0.24 (top), Y=0.30 (middle) and Y=0.38 (bottom).  
The number of stars located inside the instability strip are labeled. 
Right -- Comparison between the period distribution of the synthetic HB 
models (black line) plotted in the left panels and the observed period 
distribution (red dashed line). The period of the first overtones were 
fundamentalized ($\log$ $P_F$=$\log$ $P_{FO}$+0.127). Predicted and 
observed period distributions were smoothed using a Gassian kernel.   
\label{fig4}}
\end{figure*}

Fig.~3 shows the period distribution of \rr stars in \omc using the sample 
adopted by \citet{dp06}. The period distribution was smoothed using a Gaussian kernel 
\citep[for more details see][]{dice10}. The comparison between theory and observations 
brings forward interesting results:
{\em i)}-- The period range covered by He-enhanced models is systematically longer 
than observed.  The discrepancy becomes even more clear if we account for the fact 
that variables with periods longer than $\approx$0.9 day are classified as 
Type II Cepheids\footnote{The long period tail of \rr stars in \omc includes seven 
objects with periods ranging from $\sim$0.81 to $\sim$0.87 day and with mean visual 
magnitudes ranging from 14.32 (V31) to 14.56 (V78) mag. These objects are well separated 
from Type II Cepheids, since there is a group of six variables with periods ranging 
from $\sim$0.97 to $\sim$2.27 day and mean visual magnitudes that are at least half 
magnitude brighter than long period \rr stars. Their mean visual magnitude range 
from 13.32 (V134) to 13.98 (V2, Kaluzny et al. 1997). The sample of Type II Cepheids 
in \omc also includes three Type II Cepheids with periods ranging from $\sim$4.27 to 
29.35 day (Kaluzny et al. 1997; Matsunaga et al. 2006).}. 
{\em ii)}-- The He-enhanced models do predict FO periods that are systematically 
longer than observed. Note that the minimum period for FO pulsators based on canonical 
models agree quite well with observed values in GCs \citep{c00,dmc04}.   
{\em iii)}-- Metal-poor (top) and metal-intermediate (bottom) He-enhanced models show the same 
behavior. 
{\em iv)}-- The period range covered by canonical \rr models agrees quite well with the 
observed one.

The anonymous referee suggested to provide a more detailed comparison between theory and observations 
using synthetic HB models. To accomplish this goal we computed a series of synthetic HB 
using the same theoretical framework adopted by Percival et al. (2009). Note that the main 
aim  of this approach is to constrain the impact that different HB sub-populations characterized 
by different He contents have on the period distribution of RR Lyrae stars. We adopted two 
different  assumptions concerning the mass distribution to populate the synthetic HBs. 
In particular, we adopted a linear sampling and a Gaussian sampling. In the former case 
we distributed --for each chemical composition-- 4,000 artificial stars according to the 
evolutionary lifetimes predicted by a detailed grid of HB models. Data plotted in the left 
panels of Fig.~4 show the synthetic HBs, while the right panels display the period distribution 
of the objects located between the blue edge of the first overtones and the red edge of the 
fundamental mode. In the comparison between predicted (solid line) and observed (red dashed 
line)  period distribution, we fundamentalize the period of the first overtones 
($\log$ $P_F$=$\log$ $P_{FO}$+0.127). The predicted period distributions were 
smoothed using the same approach adopted for the observed one. Data plotted in the 
top right panel indicate that a significant fraction of HB stars populating the warm 
and the hot region of the HB might have a canonical helium content.

The minimum fundamentalized period is a robust observable to constrain the 
evolutionary properties of RR Lyrae stars 
\citep[the interested reader is referred to][and references therein]{bccm95}. 
We found that the observed minimum fundamentalized period is $\sim$ 0.34 day, while the 
predicted one, assuming a canonical helium abundance is $\sim$ 0.39 day. Moreover, 
the predicted distribution shows a well defined upper limit for RR Lyrae stars at 
$P\sim$0.87 day as suggested by the observations. 
The shape of the predicted period distribution is broader than the observed one, 
but this difference is out of the aim of this 
investigation. Note that the predicted fraction of Type II Cepheids with periods  
0.95$\le$P$\le$2.30 is a factor of three smaller ($\sim$1\%) than the observed one 
($\sim$3\%). Empirical evidence indicates that only six  
Type II Cepheids with 0.95$\le$ P $\le$2.3 have been identified 
in \omcp (see Kaluzny et al. 1997). 

The synthetic HBs computed assuming the He-enhanced HB models with Y=0.30 show a 
period distribution that is systematically shifted toward longer periods when compared 
with the observed one (see the middle right panel of Fig.~4). Moreover, this period 
distribution shows, at odds with observations, two well separated peaks. 
The comparison with the period distribution based on HB models with Y=0.38 
is more difficult, since these structures do not produce RR Lyrae stars, 
but only Type II Cepheids.  However, the predicted Type II Cepheids attain 
visual magnitudes that are brighter (V$\le$13.2$\pm$0.1 mag, see the bottom 
right panel of Fig.~4) than the observed ones (see Kaluzny et al. 1997). 
In passing we note that the crossing time of the He-enhanced models, at fixed
 metal content (Z=0.001), is on average 10\% longer (Bono 2010) than for models 
with canonical He-content. This means that the probability to produce Type II Cepheids 
for He-enhanced structures is higher than for canonical ones. 
     
\begin{figure*}
\includegraphics[height=0.5\textheight,width=0.9\textwidth]{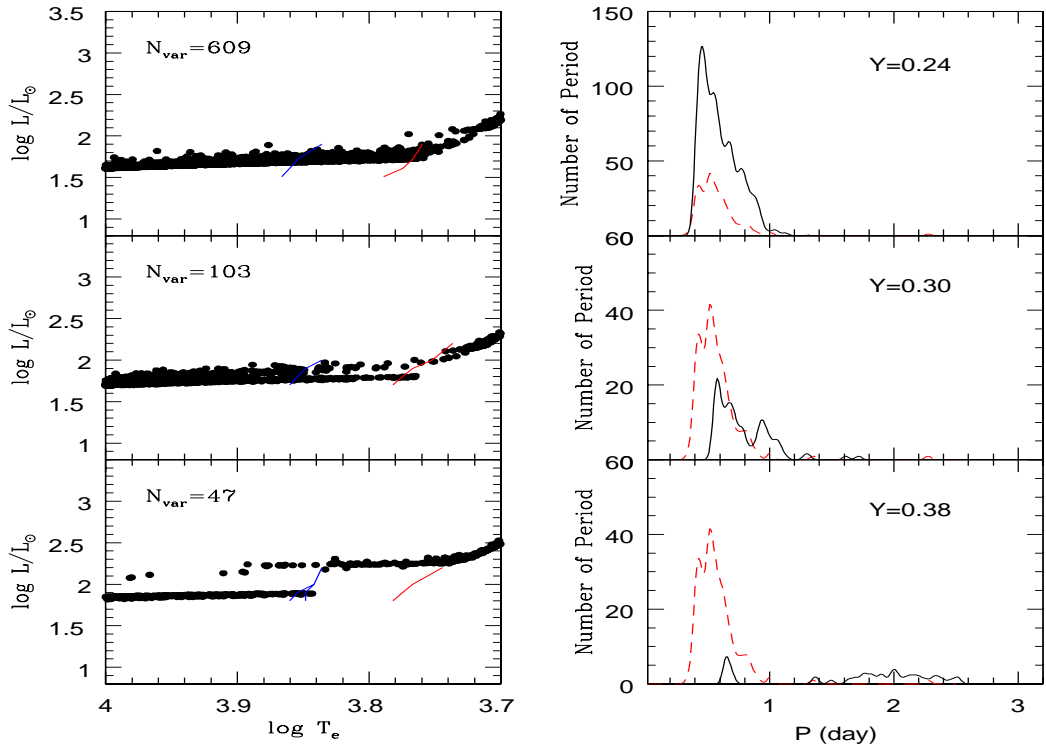}
\caption{Same as Fig.~4, but the synthetic HB models were constructed 
assuming a Gaussian sampling in mass with different mean masses and the 
same dispersion. See text for more details.\label{fig5}}
\end{figure*}

The synthetic HB models based on the Gaussian sampling in mass, were constructed 
adopting for each chemical composition a slightly different value of the mean mass 
(M/M$_\odot$=0.62 [Y=0.246], 0.61 [Y=0.30], 0.60 [Y=0.38]) populating the HB 
and the same dispersion in mass ($\sigma$=0.02 M$_\odot$). 
Data plotted in Fig.~5 show that the period distribution based on synthetic 
HB models with Gaussian sampling in mass is  very similar to the models with 
linear sampling. The minimum fundamentalized period of the canonical models 
is $\sim$ 0.39 day, and the fraction of Type II Cepheids with periods  
0.95 $\le$ $P$ $\le$2.30 is similar (2\%) to the observed one. 
Moreover, the predicted period distribution shows a well defined peak 
similar to the observed one. The period distributions based on He-enhanced 
models show discrepancies similar to synthetic HBs computed assuming a 
linear sampling in mass.     

\begin{figure}
\includegraphics[height=0.5\textheight,width=0.5\textwidth]{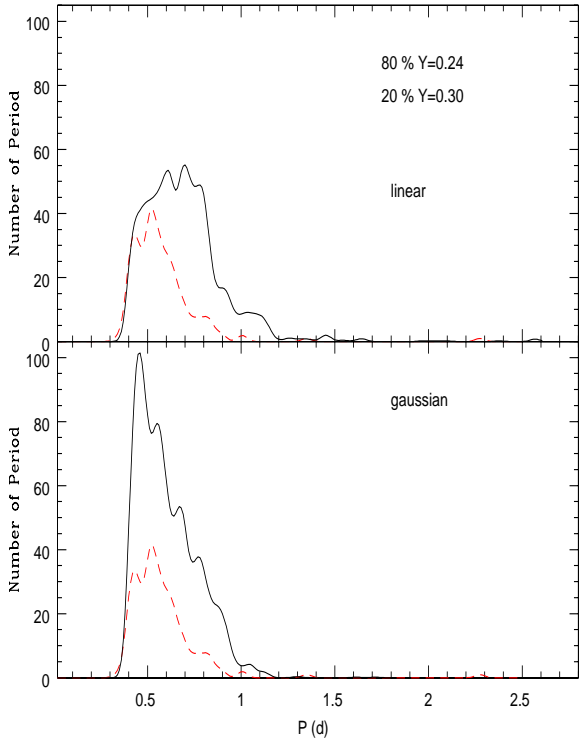}
\caption{Top -- Comparison between the observed period distribution 
and the predicted cumulative period distribution based on the linear 
sampling in mass (see Fig.~4). The latter was estimated assuming the 
80\% of the canonical and the 20\% of the helium-enhanced (Y=0.30) 
period distribution. Bottom -- Same as the top, but the predicted 
cumulative distribution is based on the Gaussian sampling 
in mass (see Fig.~5).
\label{fig6}}
\end{figure}

However, the above conclusions rely on the assumption that warm and hot HB stars 
are only made by HB structures either with a canonical or with a He-enhanced 
compositions. Recent investigations concerning the fraction of second generations,  
He-enhanced stars in GCs indicate that it might be of the order of 50\% or even more 
(D'Antona \& Caloi 2008). For $\omega$ Cen it was suggested that the fraction of 
stars with a moderate increase in He content (Y$\le$ 0.38) should be of the order 
of 20\% (Norris 2004; Lee et al. 2005). To account for the possible 
mix of RR Lyrae stars with different helium contents Fig.~6 shows the cumulative 
period distribution computed assuming the 80\% of the period distribution based 
on canonical HB models and the 20\% of the period distribution based on helium-enhanced 
HB models (Y=0.30). Note that we did not apply any normalization between the two 
period distributions. The cumulative period distribution based on the uniform mass 
distribution  shows a minimum fundamentalized period that is, as expected, very 
similar to the period distribution based on canonical HB models. However, the 
fraction of Type II Cepheids with periods 0.95$\le$ $P$ $\le$2.30 is almost  
a factor of three larger (8\%) than the observed one. The outcome is similar 
for the cumulative period distribution based on the Gaussian sampling, but 
the same fraction is almost a factor of two larger (5\%) than the observed 
one (see the bottom panel of Fig.~6).    

\begin{figure}
\includegraphics[height=0.5\textheight,width=0.5\textwidth]{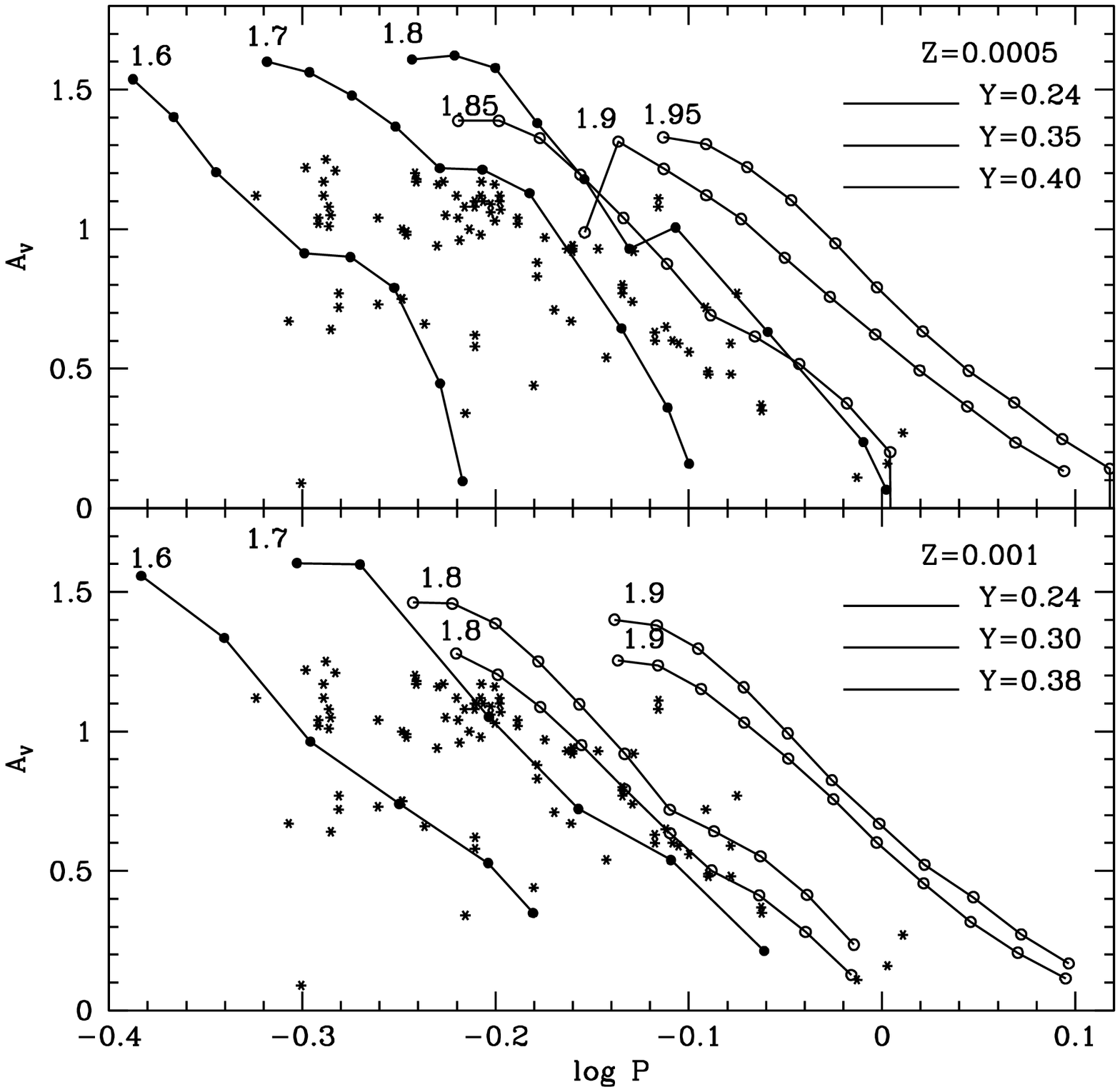}
\caption{Top -- Comparison between observed \citep{Kal97} and predicted V-band 
amplitudes versus the logarithmic period for metal-poor ($Z=0.0005$) \rr  
models. The adopted He abundances and luminosity levels are labelled.
Bottom -- Same as the top, but for metal-intermediate ($Z=0.001$) \rr  
models.\label{fig7}}
\end{figure}

The above findings indicate that the fraction of RR Lyrae stars with Y=0.30
cannot be larger than 20\%. The occurrence of He-enhanced structures causes a 
steady increase in the typical period of RR Lyrae stars and an increase in 
the fraction of Type II Cepheids with periods between 0.95 and 2.30 day that 
is not supported by observations. These results are minimally affected by 
the sampling in mass adopted to compute the synthetic HB models. 
We cannot reach firm conclusions concerning the fraction of HB stars with 
Y=0.38, since these structures only produce Type II Cepheids. However, 
the predicted visual magnitudes for these objects are systematically 
brighter than observed.

To further constrain the impact of the He abundance on the pulsation properties 
of \rr stars we also compared observed \citep{Kal97} and predicted V-band amplitudes
for FU RR Lyrae. Note that luminosity amplitudes are also independent of cluster 
distance and reddening corrections.  
Bolometric light curves were transformed into the observational plane using the bolometric 
corrections and the Color-Temperature transformations provided by \citep{cast97a,cast97b}.  
Fig.~7 shows the comparison between observed and predicted V-band amplitudes for canonical
(black lines) and He-enhanced (colored lines) \rr models. Data plotted in this figure 
indicate, once again, that the He content marginally affects the pulsation behavior of 
RR Lyrae. The top panel shows that the set of metal-poor models with an increase of 
50\% in He, when compared with canonical models and with similar luminosities 
($\log L/L_\odot \approx$1.85 vs 1.80), display very similar amplitudes and 
cover very similar periods. The same outcome applies to the two sets of 
He-enhanced, metal-intermediate \rr models (bottom panel).

Data plotted in this figure indicate that He-enhanced models only account for the upper 
envelope of the observed distribution. An increase either in the luminosity level or in 
the He content increases the discrepancy between theory and observations. The canonical 
\rr models are in fair agreement with observations. However, in dealing with luminosity 
amplitudes we need to keep in mind two limits.  
{\em i)}-- The theoretical Bailey diagram is affected by uncertainties on the mixing length 
parameter \citep[see e.g.][]{m03}.Current pulsation models were constructed by 
assuming a mixing length parameter $\alpha= 1.5$. In the mixing length formalism, 
a larger  $\alpha$ value means larger efficiency of convective motions, and in turn, 
smaller pulsation amplitudes. However, we have verified (Marconi et al. 2011, in preparation) 
that by increasing the mixing length parameter from 1.5 to 2.0,  the He-enhanced models 
do not fit the distribution of the observed amplitudes.
Several numerical experiments performed using our theoretical framework show that 
smaller $\alpha$ values, as the ones recently suggested by asteroseismological 
results\citep[][and references therein]{piau11}, would predict pulsation observables for 
RR Lyrae variables that are at odds with observations. This outcome applies not only to 
the Bailey diagram, but also to the modal stability and the topology of the instability 
strip. 
{\em ii)}-- Recent space (COROT, Chadid et al. 2010) and ground-based \citep{kun10} 
observations indicate that the fraction of \rr stars affected by the Blazkho 
phenomenon is higher than previously estimated ($\approx$50\%, \citet{ben10}. 
This effect causes a modulation of both amplitudes and phase and unfortunately, 
we still lack a complete census of Blazkho \rr  in \omcp.     

\section{Conclusions and final remarks}

To constrain the possible occurrence of He-enriched \rr stars we adopted two observables 
--periods, V-band amplitudes-- that are independent of cluster distance and reddening. 
The comparison between theory and observations indicates that the predicted periods and 
amplitudes of \rr stars are marginally affected by the He content. The key parameter 
causing the difference between canonical and He-enhanced observables is the luminosity. 
We found that the period range of He-enhanced \rr models is systematically longer 
than observed. These findings apply to metal-poor and metal-intermediate He-enhanced 
models.  
The results are the same if we take into account the luminosity amplitude. However, 
predicted amplitudes are less robust than the pulsation periods, since they depend 
on the adopted mixing length parameter. 
Moreover, the amplitude might also be affected by an observational bias, since 
we still lack a complete census of \omc \rr affected by the Blazkho 
phenomenon.  

Taken at face value the above results indicate that \omc \rr do not 
show a clear evidence of He-enrichment when moving from metal-poor to metal-intermediate 
stars. However we cannot exclude, as originally suggested by Sollima et al. (2006), the 
probable occurrence of two metal-intermediate sub-populations. The former one with 
canonical He content evolving into the \rr IS and the latter one with 
a He-enhanced abundance evolving into the hot and the EHB region. 
Note that \omc hosts more than 2,100 hot HB stars, i.e. roughly the
65\% of the entire population of HB stars (see Table 2 in Castellani et al. 2007). 

To constrain the occurrence of He-enhanced RR Lyrae stars we also computed a series
of synthetic HB models using canonical and He-enhanced HB models and two different 
assumptions concerning the sampling in mass (linear, Gaussian). Current models further 
confirm the sensitivity of the minimum fundamentalized period to constrain the He 
content of cluster RR Lyrae.  The period distribution based on canonical HB models 
agree quite well with observations concerning the minimum period (0.39 vs 0.34), 
the upper limit to the period distribution of RR Lyrae stars (P$\sim$0.87 day) 
and the fraction (2\% vs 3 \%) of Type II Cepheids (0.95$\le$P$\le$2.30 day). 
These findings minimally depend on the adopted sampling in mass.  

The period distribution based on He-enhanced HB models is either systematically 
shifted towards longer periods (Y=0.30) or produce Type II Cepheids that are 
systematically brighter than observed (Y=0.38). 
To further constrain the impact on the period distribution of RR Lyrae with different 
He contents we computed the period distribution of a sample made with the 80\% of 
canonical and the 20\% of RR Lyrae stars with moderate He-enhancement (Y=0.30). 
We found that the fraction of Type II Cepheids predicted by the synthetic 
HB models computed by assuming a Gaussian sampling in mass is almost a factor 
of two larger than observed (5\%  vs 3\%). This indicates that the fraction of 
He-enhanced structures in \omc cannot be larger than 20\%.  
 
More quantitative constraints concerning the individual He abundances of 
RR Lyrae stars in \omc requires accurate estimates of the 
A-parameter\footnote{The A-parameter is a diagnostic to constrain the 
mass-luminosity ratio of individual RR Lyrae stars. It is defined as: 
A$=\log{L/L_{\odot}}-0.81 \log{M/M_{\odot}}$, where $L$ and $M$ are the 
luminosity and the mass of the pulsator.} \citep{c83,sand00}  
using homogeneous grids of pulsation and evolutionary models constructed 
assuming a broad range of helium abundances 
(Caputo et al. 2011, in preparation).   

Finally, in a recent investigation \citep{du11} detected in five out of twelve 
RGs in \omcp, with similar magnitudes and colors, the chromospheric 
HeI line at 10830\AA. The strength of the He line seems to be correlated with 
Al and Na abundances rather than with iron. However, a more detailed non-LTE 
analysis of the He abundance is required before firm conclusions concerning 
the occurrence of a spread in helium can be safely established.

\acknowledgments
It is a pleasure to thank G. Iannicola and I. Ferraro for many useful 
suggestions concerning the Gaussian smoothing of the period distribution. 
We thank an anonymous referee for his/her comments and suggestions 
that improved the content and the readability of the manuscript.
The authors acknowledge financial support through the project PRIN MIUR 2007 
(P.I.: G.P. Piotto) and from PRIN INAF 2009 (P.I.: R. Gratton).

{}


\begin{thebibliography}{}

\bibitem[Alexander \& Ferguson(1994)]{af94} Alexander, D. R., Ferguson, J. W. 1994, ApJ, 437, 879

\bibitem[Anderson(2002)]{ande02} Anderson, J. 2002, ASPC, 265, 87 

\bibitem[Bedin et al. (2004)]{be04} Bedin, L. R. et al. 2004, ApJ, 605, 125

\bibitem[\protect\citeauthoryear{{Bellini} et al.}{{Bellini} et al.}{2010}]{bel10}
Bellini, A., Bedin, L. R., Piotto, G., Milone, A. P., Marino, A. F., Villanova, S. 2010, AJ, 140, 631

\bibitem[Benko et al.(2010)]{ben10}  Benko, J. M., Kolenberg, K., Szab\'o, R. et al. 2010, MNRAS, 409, 1585



\bibitem[Bergbusch \& Stetson(2009)]{berst09} Bergbusch, P. A., Stetson, P. B. 2009, AJ, 138, 1455

\bibitem[Bono (2010)]{bo10} Bono, G. 2010, MmSAI, 81, 863

  
\bibitem[Bono, Marconi, \& Stellingwerf(1999)]{bms99} Bono, G., Marconi, M., Stellingwerf, R.F. 1999, \apjs, 122, 167                                                                                                      
\bibitem[Bono et al. (1995)]{bccm95} Bono, G., Caputo, F., Castellani, V., Marconi, M. 1995, \apjl, 448, 115

\bibitem[Bono et al.(2003)]{b03}  Bono, G., Caputo, F., Castellani, V., Marconi, M., Storm, J., Degl'Innocenti, S. 2003, MNRAS, 344, 1097

\bibitem[Bono et al.(2008)]{bono08} Bono, G., Stetson, P. B., Sanna, N. et al. 2008, \apj, 686, 87

\bibitem[Bono \& Stellingwerf(1994)]{bs94} Bono, G., Stellingwerf, R.F. 1994, ApJS, 93, 233                   

\bibitem[Brown et al. (2010)]{Brown10} Brown, T. M., Sweigart, A. V., Lanz, T., Smith, E.  et al. 2010, \apj, 718, 1332

\bibitem[Calamida et al.(2007)]{cala07} Calamida, A., Bono, G., Stetson, P. B.  et al. 2007, \apj,  670, 400

\bibitem[Calamida et al.(2009)]{cala09} Calamida, A., Bono, G., Stetson, P. B.  et al. 2009, \apj,  706, 1277

\bibitem[Caputo et al.(1983)]{c83} Caputo F., Cayrel R., Cayrel de Strobel G., 1983, A\&A, 123, 135 

\bibitem[Caputo et al.(2000)]{c00} Caputo, F., Castellani, V., Marconi, M., Ripepi, V. 2000, MNRAS, 316, 819

\bibitem[Cassisi et al.(2008)]{cass08}Cassisi, S., Salaris, M., Pietrinferni, A. et al. 2008, ApJ, 672,L115


\bibitem[Cassisi et al.(2009)]{cas09} Cassisi, S., Salaris, M., Anderson, J., et al. 2009, \apj, 702, 1530

\bibitem[Castellani et al.(2007)]{caste07} Castellani, V., Calamida, A., Bono, G. et al. 2007, ApJ, 663, 1021

\bibitem[Castelli, Gratton \& Kurucz(1997a)]{cast97a} Castelli, F., Gratton, R. G., \& Kurucz, R. L. 1997a, A\&A, 318, 841 

\bibitem[Castelli, Gratton \& Kurucz(1997b)]{cast97b} Castelli, F., Gratton, R. G., \& Kurucz, R. L. 1997b, A\&A, 324, 432 

\bibitem[Chadid et al. (2010)]{cha10} Chadid, M., Benko, J. M., Szab\'o, R. et al. 2010, A\&A, 510, 39


\bibitem[D'Antona, Caloi \& Ventura (2010)]{dan10} D'Antona, F., Caloi, V., Ventura, P. 2010, MNRAS, 405, 2295

\bibitem[D'Antona \& Caloi(2008)]{dant08} D'Antona, F. \& Caloi, V. 2008, \mnras, 390, 693

\bibitem[Del Principe et al. (2006)]{dp06} Del Principe et al. 2006, ApJ, 652, 362

\bibitem[Di Cecco et al.(2010)]{dice10} Di Cecco, A., Bono, G., Stetson, P. B. et al. 2010, ApJ, 712, 527 

\bibitem[Di Criscienzo, Marconi \& Caputo (2004)]{dmc04} Di Criscienzo, M., Marconi, M., Caputo, F. 2004, ApJ, 612, 1092


\bibitem[Di Criscienzo et al. (2010)]{dicri10} Di Criscienzo, M., Ventura, P., D'Antona, F., 
Milone, A., \& Piotto, G. 2010, MNRAS, 408, 999  

\bibitem[Dupree, Strader \& Smith (2011)]{du11} Dupree, A. K., Strader, J., Smith, G. H. 2011, ApJ, 728, 155

\bibitem[Gratton, Sneden \& Carretta (2004)]{grat04} Gratton, R., Sneden, C., \& Carretta, E. 2004, \aap, 42, 385

\bibitem[Iglesias \& Rogers (1996)]{ir96} Iglesias, C., Rogers, F. J. 1996, ApJ, 464, 943

\bibitem[Kaluzny et al. (1997)]{Kal97} Kaluzny, J., Kubiak, M., Szymanski, M., Udalski, A., Krzeminski, W., Mateo, M. 1997, A\&AS, 125, 343

\bibitem[Kaluzny et al. (2004)]{Kal04} Kaluzny, J., Olech, A., Thompson, I. B., Pych, W., Krzemi\'nski, W., Schwarzenberg-Czerny, A. 2004, A\&A, 424, 1101

\bibitem[Kraft (1994)]{Kraft94} Kraft, R. P. 1994, PASP, 106, 553

\bibitem[Kunder, Chaboyer, \& Layden (2010)]{kun10} Kunder, A., Chaboyer, B., Layden, A. 2010, AJ, 139, 415

\bibitem[Lee et al. (2005)]{lee05} Lee, Y.-W., Joo, S.-J., Han, S.-I. et al. 2005, ApJ, 621, 57

\bibitem[Lee et al. (2009b)]{lee09} Lee, J.-W., Kang, Y.-W., Lee, J., Lee, Y.-W. 2009b, Nature, 462, 480

\bibitem[Marconi et al.(2003)]{m03} Marconi, M., Caputo, F., Di Criscienzo, M., Castellani, M. 2003, ApJ, 596, 299                                        

\bibitem[Miller Bertolami et al. (2008)]{Miller08}  Miller Bertolami, M. M., Althaus, L. G., Unglaub, K., \& Weiss, A. 2008, \aap, 491, 253

\bibitem[Milone et al.(2008)]{milo08} Milone, A. P., Bedin, L. R., Piotto, G. et al. 2008, ApJ, 673, 241

\bibitem[Moehler et al.(2010)]{moeh10} Moehler, S. Dreizler, S., Lanz, T., Bono, G., 
Sweigart, A. V., Calamida, A., Nonino, M. 2011, A\&A, 526, 136  

\bibitem[Nataf et al.(2011)]{nata11} Nataf, D. M., Gould, A., Pinsonneault, M. H., Stetson, P. B. 2011, submitted ApJ, arXiv1102.3916  

\bibitem[Norris (2004)]{norr04} Norris, J. E. 2004, ApJ, 612, 25

\bibitem[Osborn (1971)]{Osborn71} Osborn, W. 1971, The Observatory, 91, 223

\bibitem[Piau et al.(2011)]{piau11} Piau, L., Kervella, P., Dib, S., Hauschildt, P. 2011, A\&A, 526, 100

\bibitem[Pietrinferni et al.(2006)]{piet06} Pietrinferni, A., Cassisi, S., Salaris, M., Castelli, F. 2006, ApJ, 642, 797 

\bibitem[Pilachowski, Sneden \& Wallerstein (1983)]{pila83} Pilachowski, C. A., Sneden, C., \& Wallerstein, G. 1983, APJS, 52, 241

\bibitem[Piotto et al. (2005)]{pio05} Piotto, G. et al. 2005, ApJ, 621, 777


\bibitem[Piotto et al. (2007)]{piot07} Piotto, G. et al. 2007, ApJ, 661, 53

\bibitem[Sandquist (2000)]{sand00} Sandquist, E. L. 2000, MNRAS, 313, 571  

\bibitem[Smith (1987)]{smith87} Smith, G. H. 1987, PASP, 99, 67l

\bibitem[Sollima et al. (2006)]{s06} Sollima, A. et al. 2006, ApJ, 640, 43

\bibitem[Smolec \& Moskalik (2010)]{smolec10} Smolec, R., Moskalik, P. 2010, A\&A, 524, 40

\bibitem[Siegel et al.(2007)]{sieg07} Siegel, M. H., et al. 2007, \apj, 667, L57

\bibitem[Stellingwerf(1982)]{s82} Stellingwerf, R. F. 1982, ApJ, 262, 330

\bibitem[Weldrake, Sackett, \& Bridges(2007)]{Wel07} Weldrake, David T. F., Sackett, Penny D., Bridges, Terry J. 2007, AJ, 133, 1447


\end{thebibliography}
\end{document}